\documentclass{iopart}
\usepackage{graphicx}
\usepackage{bm}

\newcommand{\hG}{\hat{G}}
\newcommand{\hH}{\hat{H}}
\newcommand{\bE}{\bar{E}}
\newcommand{\cE}{\bar{\cal E}}

\begin{document}
\title[Green functions of electrons in graphene in a magnetic field]
      {Green functions of electrons in monolayer and bilayer graphene in a magnetic field}
\date{\today}
\author{Tomasz M Rusin\dag\ and Wlodek Zawadzki*}
\address{ \dag Orange Customer Service sp. z o. o., ul. Twarda 18, 00-105 Warsaw, Poland\\
         *Institute of Physics, Polish Academy of Sciences, 02-668 Warsaw, Poland}
 \ead{Tomasz.Rusin@centertel.pl}

\pacs{71.70.Di,73.22.Pr}
\submitto{\JPA}

\begin{abstract}
Closed expressions for the Green functions of the stationary two-dimensional two-component
Schrodinger equation for an electron moving in monolayer and bilayer graphene in the presence of a
magnetic field are obtained in terms of the Whittaker functions.
\end{abstract}

\maketitle
Sometime ago Dodonov {\it et al.}~\cite{Dodonov1975} calculated
the stationary Green function of a free 2D electron in a homogenous magnetic field and
obtained analytical results in terms of the Whittaker functions.
Similar problems were recently investigated for low-dimensional systems
\cite{Gusynin1995,Gorbar2002,Murguia2010}, where the Green functions were obtained as
infinite sums of Laguerre polynomials.
Recently, problems of propagators were analyzed theoretically for electrons in monolayer graphene
\cite{Horing2009,Champel2010}. Horing and Liu~\cite{Horing2009} obtained a propagator as an infinite sum
and, alternatively, as the second solution of the Bessel wave equation.
A closed form of the propagator in monolayer graphene in terms of the confluent hypergeometric function
was obtained by Piatkovskii and Gusynin~\cite{Piatkovskii2010}.
To our knowledge, the expression for the Green function in terms of the Whittaker functions
has not been published in the literature. The problem of Green function
for bilayer graphene in a magnetic field has not been analyzed earlier.

One should add here that a somewhat related problem was tackled many years ago
by Schwinger~\cite{Schwinger1951}, who considered the Green function for a Dirac electron
in a vacuum in external fields. However, the solution of this problem was found employing the
proper-time formalism in which one can directly use the gauge-invariant fields rather
than potentials. Also, our problem is distinctly different since the Dirac electron
is characterized by a rest mass~$m_0$ and an energy gap~$2m_0c^2$, while the band structure
of monolayer graphene has a vanishing ``rest mass'' and a vanishing energy gap. In turn,
in bilayer graphene there is no gap and the energy bands are parabolic.

The aim of the present work is to derive an analytical closed form of the stationary
electron Green function for electrons in monolayer and bilayer
graphene in a uniform magnetic field. The electron Green function is used
in calculating the local density of states, scattering processes, transport and disorder
properties of a material, as well as in many-body problems.

In the first step we consider the stationary Green function of a two-dimenional free electron
in a magnetic field. The spin is omitted. The two-dimensional Hamiltonian is
$\hH^e=\hat{\bm \pi}^2/(2m)$, where $\hat{\bm \pi} = \hat{\bi{p}} + e\bi{A}$
and~$\bi{A}$ is the vector potential. In the Landau gauge $\bi{A} =(-By,0)$,
the eigenstates of~$\hH^e$ are $\psi_{nk_x}^e({\bm \rho})=e^{ik_xx}\phi_n(\xi)/\sqrt{2\pi}$, where
$\phi_n(\xi) = {\sqrt{L}C_n}{\rm H}_{n}(\xi)e^{-1/2\xi^2}$,
${\rm H}_{n}(\xi)$ are the Hermite polynomials, $C_n=\sqrt{2^n n!\sqrt{\pi}}$,
the magnetic radius is $L = \sqrt{\hbar/eB}$ and $\xi=y/L-k_xL$. The energy levels are
$E_{n}=\hbar \omega_c (n+1/2)$ with $\omega_c=eB/m$. The Green function is by definition
\begin{equation} \label{eGDef}
\hG_e({\bm \rho}, {\bm \rho}', E) = \sum_{n}\!\int_{-\infty}^{\infty}\!\!
  \frac{e^{ik_x(x-x')}\phi_{n}(\xi) \phi_{n}(\xi')^*}
  {2\pi[\hbar \omega_c(n+1/2) - E]} dk_x,
\end{equation}
where ${\bm \rho}=(x,y)$. To integrate over~$k_x$ we use the identity~\cite{GradshteinBook}
\begin{equation} \label{eHmHn}
\int_{-\infty}^{\infty}\!\!\!\! e^{-x^2}{\rm H}_m(x+y) {\rm H}_n(x+z)dx = 2^n\sqrt{\pi}m!z^{n'}{\rm L}_m^{n'}(-2yz),
\end{equation}
where $n'=n-m$ and $m\leq n$. This gives
\begin{equation} \label{eGLaguerre}
\hG_e({\bm \rho}, {\bm \rho}', E) = \frac{e^{-r^2/2+i\chi}}{2\pi\hbar \omega_c L^2}
   \sum_{n=0}^{\infty} \frac{L_n^0(r^2)}{n+1/2-\cE},
\end{equation}
where $\cE=E/(\hbar\omega_c)$, $L_n^{\alpha}(r^2)$ are the associated Laguerre polynomials,
$r^2 = ({\bm \rho} - {\bm \rho}')^2/(2L^2)$ and $\chi=(x-x')(y+y')/2L^2$ is the gauge-dependent phase factor.
The summation over~$n$ in~(\ref{eGLaguerre}) can be performed with the use of
formula 6.12.4 in~\cite{ErdelyiBook}
\begin{equation} \label{ePsi}
 t^{-\alpha} \sum_{n=0}^{\infty} \frac{L_n^{-\alpha}(t)}{n+a-\alpha} = \Gamma(a-\alpha)\Psi(a,\alpha+1;t),
\end{equation}
where $\Psi(a,c;t)$ is the second solution of the confluent hypergeometric equation~\cite{ErdelyiBook}.
The series in~(\ref{ePsi}) converges for~$t>0$ and~$\alpha>-1/2$.
There is (see formula 6.9.4 in~\cite{ErdelyiBook})
\begin{equation} \label{ePsiW}
\Psi(a,c;t)=e^{t/2}t^{-1/2-\mu}W_{\kappa,\mu}(t),
\end{equation}
where $W_{\kappa,\mu}(t)$ is the Whittaker function, $\kappa=c/2-a$ and $\mu=c/2-1/2$.
On combining equations~(\ref{eGLaguerre}), (\ref{ePsi}) and~(\ref{ePsiW}), setting
$\alpha=0$, $t=r^2$ and $a=1/2-\cE$ we obtain: $c=1$, $\mu=0$ and $\kappa=\cE$, so that the free electron
Green function is
\begin{equation} \label{eGDodonov}
\hG_e({\bm \rho},{\bm \rho}', E) = \frac{e^{i\chi}}{2\pi\hbar\omega_c L^2|r| }\Gamma(1/2-\cE)W_{\cE,0}(r^2).
\end{equation}
The same result was obtained by Dodonov {\it et al.}~\cite{Dodonov1975}
using the Laplace transform of the time-dependent electron
Green function $\hG_e({\bm \rho},{\bm \rho}', \beta)$.
Comparing~(\ref{eGLaguerre}) and~(\ref{eGDodonov}) we have an important
auxiliary result for free electrons employed in the calculations below
\begin{equation} \label{eBaseId}
e^{-r^2/2} \sum_{n=0}^{\infty} \frac{L_n^0(r^2)}{n+1/2-\cE} =
 \frac{1}{|r| }\Gamma(1/2-\cE)W_{\cE,0}(r^2).
\end{equation}

Now we turn to the main subject of our work. The Hamiltonian for electrons at the~$K$ point of the Brillouin zone
in monolayer graphene in a magnetic field~$B$ is $\hH^M = u\hat{\sigma}_x \hat{\pi}_x + u\hat{\sigma}_y \hat{\pi}_y$,
where $u\simeq 1 \times 10^8$ cm/s is the band electron velocity
and~$\hat{\sigma}_x$,~$\hat{\sigma}_y$ are
the Pauli matrices~\cite{Wallace1947,Semenoff1984}. In the Landau gauge, the eigenstates of~$\hH^M$ are
\begin{equation} \label{ML_psi_n}
\psi_{nk_xs}^M({\bm \rho}) = \frac{e^{ik_xx}}{\sqrt{2\pi(2-\delta_{n,0})}} \left(\begin{array}{c}
   -s\phi_{n-1}(\xi) \\ \phi_n(\xi) \end{array}\right),
\end{equation}
where $s=\pm 1$ and $\phi_n(\xi)$ are defined above. The energy levels are
$E_{ns}=s \hbar \omega \sqrt{n}$ with $\omega=\sqrt{2}u/L$.

The stationary Green function of the Hamiltonian~$\hH^M$ is a $2\times 2$ matrix
$\hG^M=\left(\begin{array}{cc} \hG_{11}^M & -\hG_{10}^M \\ -\hG_{01}^M & \hG_{00}^M \end{array}\right)$,
where
\begin{equation}
 G_{\sigma\sigma}^M({\bm \rho},{\bm \rho}', E) = \sum_{n,s}\!\int_{-\infty}^{\infty}\!\!
  \frac{e^{ik_x(x-x')}\phi_{n-\sigma}(\xi) \phi_{n-\sigma}(\xi')^*}
  {2\pi(2-\delta_{n,0})(s\hbar \omega \sqrt{n} -E)} dk_x,
\end{equation}
\begin{equation}
 G_{\sigma,\sigma'}^M({\bm \rho},{\bm \rho}', E) = \sum_{n,s}\!\int_{-\infty}^{\infty}\!\!
  \frac{se^{ik_x(x-x')}\phi_{n-\sigma}(\xi) \phi_{n-\sigma'}(\xi')^*}
  {4\pi(s\hbar \omega \sqrt{n} -E)} dk_x,
\end{equation}
in which $\sigma,\sigma'=0,1$ and $\sigma \neq \sigma'$.
Performing the summation over~$s$ and integration over~$k_x$ with the use of
identity~(\ref{eHmHn}) we obtain
\begin{eqnarray} \label{ML_GDiag}
 \hG_{\sigma\sigma}^M({\bm \rho},{\bm \rho}', E) = \frac{\bE e^{-r^2/2+i\chi}}{2\pi\hbar \omega L^2}
   \sum_{n=0}^{\infty} \frac{L_n^0(r^2)}{n+\sigma-\bE^2},\\
\label{ML_GOffDiag}
 \hG_{\sigma,1-\sigma}^M({\bm \rho},{\bm \rho}', E) =
       r_{\sigma,1-\sigma}\frac{e^{-r^2/2+i\chi}}{2\pi\hbar \omega L^2}
   \sum_{n=1}^{\infty} \frac{L_{n-1}^1(r^2)}{n-\bE^2}, \ \ \
\end{eqnarray}
where $\bE=E/(\hbar\omega)$,
$r_{1,0}=[(y'-y)-i(x-x')]/(L\sqrt{2})$ and $r_{0,1}=[(y-y')-i(x-x')]/(L\sqrt{2})$.

Now we make use of~(\ref{eBaseId}).
Making the substitutions $\cE \rightarrow \bE^2+1/2$ for~$\sigma=0$ and
$\cE \rightarrow \bE^2-1/2$ for~$\sigma=1$, the
sum over~$n$ in~(\ref{ML_GDiag}) can be converted into the sum appearing in~(\ref{eBaseId}).
Thus the diagonal terms of $\hG^M({\bm \rho},{\bm \rho}', E)$ are
\begin{eqnarray}
\label{ML_G_00_fin}
\hG_{00}^M({\bm \rho},{\bm \rho}', E) &=& \frac{\bE e^{i\chi}}{2\pi\hbar\omega L^2|r|} \Gamma(-\bE^2)
       W_{\bE^2+\frac{1}{2},0}(r^2), \\
\label{ML_G_11_fin}
\hG_{11}^M({\bm \rho},{\bm \rho}', E) &=& \frac{\bE e^{i\chi}}{2\pi\hbar\omega L^2|r|} \Gamma(1-\bE^2)
            W_{\bE^2-\frac{1}{2},0}(r^2). \ \ \ \
\end{eqnarray}

To calculate the off-diagonal elements of $\hG^M({\bm \rho},{\bm \rho}',E)$ we use the identity
$L_{n-1}^1(r^2)=n[(L_{n-1}^0(r^2)-L_{n}^0(r^2)]/r^2$. Putting it into~(\ref{ML_GOffDiag}) we
obtain after simple calculation
\begin{eqnarray} \label{ML_G_off_fin}
\hG_{\sigma,1-\sigma}^M ({\bm \rho},{\bm \rho}', E) = \frac{r_{\sigma,1-\sigma}\bE}{r^2} (\hG_{11}^M - \hG_{00}^M).
\end{eqnarray}
Equations~(\ref{ML_G_00_fin})-(\ref{ML_G_off_fin}) are the final results for the
stationary Green function of an electron in monolayer graphene in the presence of
a homogenous magnetic field. The poles of $\hG^M({\bm \rho},{\bm \rho}', E)$
occur for $\bE_{ns}=0,\pm 1,\pm \sqrt{2}, \pm \sqrt{3},\ldots \pm\sqrt{n}$, which is a direct consequence of
the above given Landau energies in monolayer graphene.
The residues of $\hG^M(E_{ns})$ can be obtained from~(\ref{ML_GDiag}) and~(\ref{ML_GOffDiag}).

In Figure~1 we plot gauge-independent part of the dimensionless Green
function $\hbar\omega L^2\hG_{\sigma\sigma}^M({\bm \rho},0,E)$ for electrons in
monolayer graphene in a magnetic field
for three values of~$\bE$. It is seen that for large~$r$ the Green
function $\hbar\omega L^2\hG^M({\bm \rho},0,E)$
decays exponentially. For large values of energy~$\bE$ the decay has an oscillating character.
As follows from~(\ref{ML_G_off_fin}), the off-diagonal components are given
by the difference of $\hG_{11}^M$ and $\hG_{00}^M$.
\begin{figure}
\includegraphics[width=8cm,height=8cm]{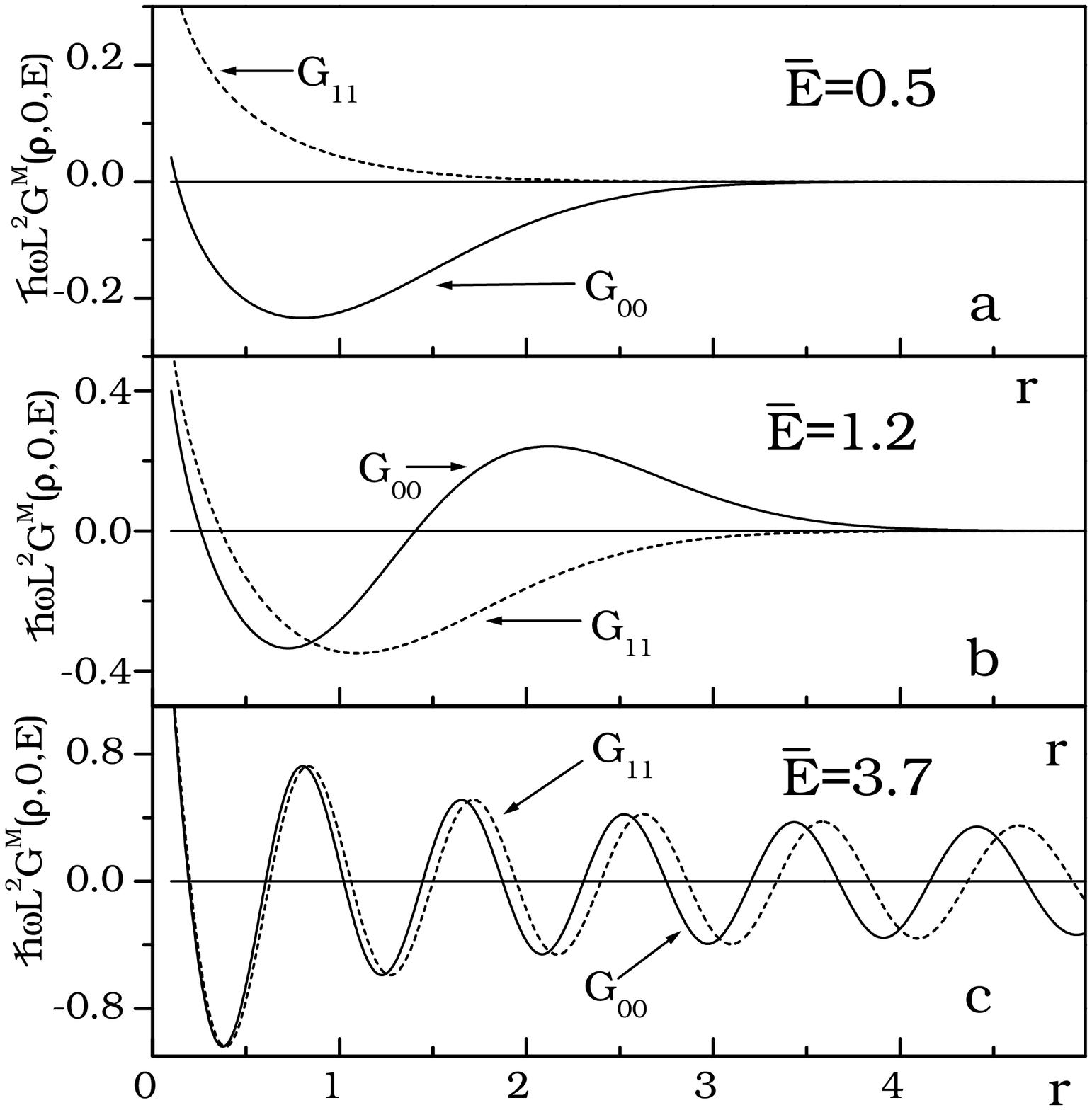}
\caption{Diagonal elements of the gauge-independent part of
         dimensionless Green function $\hbar\omega L^2\hG^M({\bm \rho},0,E)$ for monolayer graphene,
         as given in~(\ref{ML_G_00_fin})
         and~(\ref{ML_G_11_fin}) for three values of $\bE=E/(\hbar\omega)$.
         Distance is measured in $r=\sqrt{x^2+y^2}/(\sqrt{2}L)$.
         All curves are calculated with the use of expansion~(\ref{W_lambda_0}).}
\end{figure}

Expressing the Green function in terms of the Whittaker functions is useful because
the latter can be either calculated using, for example, procedures given in Mathematica,
or conveniently computed from the formula
\begin{eqnarray} \label{W_lambda_0}
W_{\lambda,0}(z) = \frac{\sqrt{z}\ e^{-z/2}}{\Gamma(1/2-\lambda)^2}
 \sum_{k=0}^{\infty}\frac{\Gamma(k-\lambda+1/2)}{(k!)^2}z^k \times \nonumber \\
  \left[2\psi(k+1)-\psi(k-\lambda+1/2) -\ln(z)\right],
\end{eqnarray}
where $\psi(z)=d\ln[\Gamma(z)]/dz$, see~\cite{GradshteinBook}.
This expansion can be obtained from the Barnes integral representation
of the Whittaker function~$W_{\cE,0}(z)$ through the calculation of residues.
In this formula, the index~$k$ labels the residues. The details of the expansion can be
found e.g. in~\cite{WangBook}. Other convenient ways to calculate
the Whittaker functions are: expansion of~$W_{\cE,0}(x)$ in terms of the Bessel functions
\cite{AbramowitzBook}, combinations of power-series expansions for small-$x$ and
and large-$x$ approximations (see~\cite{Dodonov1975,GradshteinBook}), or numerical
solutions of the Whittaker equation.

We have compared numerically~(\ref{ML_G_00_fin}),~(\ref{ML_G_11_fin}) and~(\ref{ML_G_off_fin})
with~(\ref{ML_GDiag}) and~(\ref{ML_GOffDiag})
for many random values of $0 < \bE <3$ and $0 < r< 3$, in which $W_{\lambda,0}(z)$ was calculated using
expansion~(\ref{W_lambda_0}). In addition, numerical values of $W_{\lambda,0}(z)$ were calculated using the
procedures given in Mathematica.
After truncating the summations in~(\ref{ML_GDiag}) and~(\ref{ML_GOffDiag}) at $n=1\times 10^6$ terms~(!)
we obtained only~$4$ to~$6$ significant digits of the exact results given in terms of the Whittaker
functions. We conclude that the expansion~(\ref{eBaseId}) of the Green function in terms of the Laguerre
polynomials converges quite slowly.

\begin{figure}
\includegraphics[width=8cm,height=8cm]{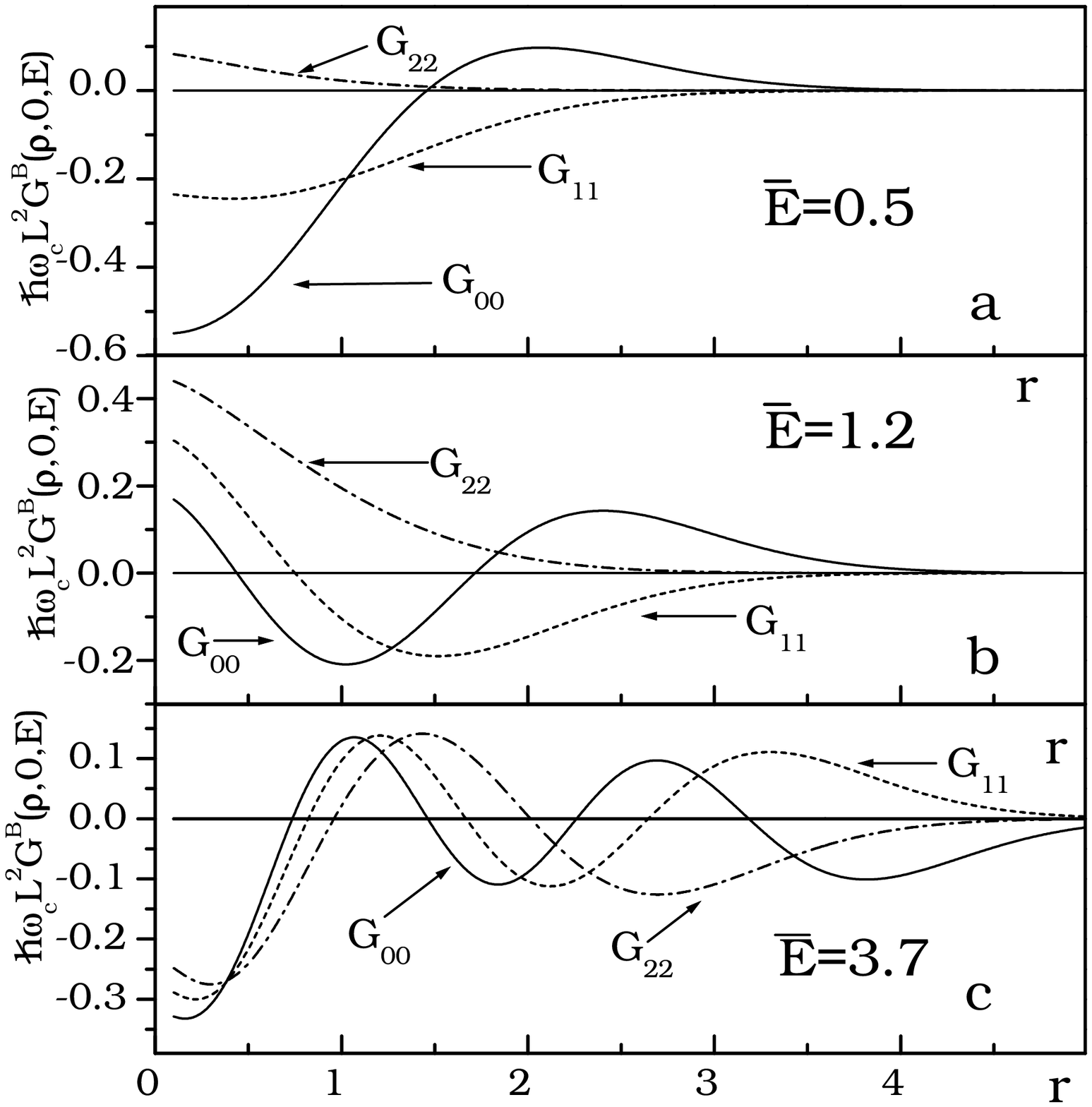}
\caption{Diagonal elements of the gauge-independent part of
         dimensionless Green function $\hbar\omega_c L^2\hG^B({\bm \rho},0,E)$ for bilayer graphene,
         as given in~(\ref{BL_G_Diag_fin}) for three values of $\bE=E/(\hbar\omega_c)$.
         Distance is measured in $r=\sqrt{x^2+y^2}/(\sqrt{2}L)$.
         All curves are calculated with the use of expansion~(\ref{W_lambda_0}).}
\end{figure}

Next, we consider electrons in bilayer graphene. At the~$K$ point of the Brillouin zone
they are described by the Hamiltonian
$\hH^B = -1/(2m^*)\left(\begin{array}{cc} 0 & (\hat{\pi}^-)^2 \\ (\hat{\pi}^+)^2 &0 \end{array}\right)$,
with $m^*=0.054\ m_e$ and $\hat{\pi}^{\pm} = \hat{\pi}_x \pm i\hat{\pi}_y$~\cite{McCann2006}.
This Hamiltonian is applicable within the energy range $|E|< 100$ meV.
To find the Green function in this case we proceed in a way similar to that described above
for monolayer graphene. In the Landau gauge the eigenstates of~$\hH^B$ are
\begin{equation} \label{BL_psi_n}
\psi_{nk_xs}^B({\bm \rho}) = \frac{e^{ik_xx}}{\sqrt{4\pi}} \left(\begin{array}{c}
   -s\phi_{n-2}(\xi) \\ \phi_n(\xi) \end{array}\right),
\end{equation}
where $s=\pm 1$ and $\phi_n(\xi)$ are defined above. The energy levels are
$E_{ns}=s \hbar \omega_c \sqrt{n(n-1)}$ with $\omega_c=eB/m^*$. The above expressions are valid
for $n\ge 2$. For $n=0,1$ the eigenstates of~$\hH^B$
are $\psi_{nk_x}^B({\bm \rho})=e^{ik_xx}/\sqrt{2\pi}\left(\begin{array}{c} 0 \\ \phi_n(\xi) \end{array}\right)$
and the corresponding energies are~$E_n=0$.

The stationary Green function of the Hamiltonian~$\hH^B$ is again a $2\times 2$ matrix
$\hG^B=\left(\begin{array}{cc} \hG_{22}^B & -\hG_{20}^B \\ -\hG_{02}^B & \hG_{00}^B \end{array}\right)$ with
\begin{equation}
 G_{\sigma\sigma}^B({\bm \rho},{\bm \rho}', E) = \sum_{n,s}\!\int_{-\infty}^{\infty}\!\!
  \frac{e^{ik_x(x-x')}\phi_{n-\sigma}(\xi) \phi_{n-\sigma}(\xi')^*}
  {4\pi(s\hbar \omega_c \sqrt{n(n-1)} -E)} dk_x,
\end{equation}
\begin{equation}
 G_{\sigma,\sigma'}^B({\bm \rho},{\bm \rho}', E) = \sum_{n,s}\!\int_{-\infty}^{\infty}\!\!
  \frac{se^{ik_x(x-x')}\phi_{n-\sigma}(\xi) \phi_{n-\sigma'}(\xi')^*}
  {4\pi(s\hbar \omega_c \sqrt{n(n-1)} -E)} dk_x,
\end{equation}
where $\sigma,\sigma'=0,2$ and $\sigma \neq \sigma'$.
Performing the summation over~$s$ and integration over~$k_x$ with the use of
identity~(\ref{eHmHn}) we obtain
\begin{equation} \label{BL_GDiag}
 \hG_{\sigma\sigma}^B({\bm \rho},{\bm \rho}', E) = \frac{\bE e^{-r^2/2+i\chi}}{2\pi\hbar \omega_c L^2}
   \sum_{n=0}^{\infty} \frac{L_n^0(r^2)}{(n+\sigma)(n+\sigma-1)-\bE^2},
\end{equation}
\begin{equation} \label{BL_GOffDiag}
 \hG_{\sigma,2-\sigma}^B({\bm \rho},{\bm \rho}', E) =
       r_{\sigma,2-\sigma}\frac{e^{-r^2/2+i\chi}}{2\pi\hbar \omega_c L^2}
   \sum_{n=2}^{\infty} \frac{L_{n-2}^2(r^2)}{n(n-1)-\bE^2}, \ \ \
\end{equation}
where $\bE=E/(\hbar\omega_c)$,
$r_{2,0}=[(y'-y)-i(x-x')]^2/(2L^2)$ and $r_{0,2}=[(y-y')-i(x-x')]^2/(2L^2)$.
To calculate $\hG_{\sigma\sigma}^B$ with the use of~(\ref{eBaseId})
we express $\hG_{\sigma\sigma}^B$ as
a combination of simple fractions $\sum_i a_i/(n-n_i)$ with suitably
chosen parameters~$\{a_i\}$ and~$\{n_i\}$. Thus we have
\begin{eqnarray} \label{BL_2Frac}
\frac{1}{(n+\sigma)(n+\sigma-1)-E^2}= \frac{1}{n_{\sigma}^+ - n_{\sigma}^-} \times \nonumber \\
 \left( \frac{1}{n+1/2-n_{\sigma}^+} - \frac{1}{n+1/2-n_{\sigma}^-} \right),
\end{eqnarray}
with
\begin{equation} \label{BL_nsigma}
n_{\sigma}^{\pm} = 1-\sigma \pm (1/2)\sqrt{1+4\bE^2}.
\end{equation}
Using~(\ref{eBaseId}) and~(\ref{BL_2Frac}) and making the substitution
$\cE \rightarrow n_{\sigma}^{\pm}$ we obtain
\begin{eqnarray}
\label{BL_G_Diag_fin}
\hG_{\sigma\sigma}^B({\bm \rho},{\bm \rho}', E)=\frac{\bE e^{i\chi}}{2\pi\hbar\omega_c L^2|r| \sqrt{1+4\bE^2}}
     \times \nonumber \\
    \left( \Gamma(1/2-n_{\sigma}^+)W_{n_{\sigma}^+,0}(r^2) - \Gamma(1/2-n_{\sigma}^-)W_{n_{\sigma}^-,0}(r^2) \right).
\end{eqnarray}
For the off-diagonal elements of~$\hG^B$ we express the associated Laguerre polynomials $L_{n-2}^2(r^2)$
by the linear combinations of $L_n^0(r^2)$ using the identity $xL_n^{a+1}(x)=(n+a+1)L_n^a(x)-(n+1)L_{n+1}^a(x)$.
This gives, in analogy to~(\ref{ML_G_off_fin}),
\begin{eqnarray} \label{BL_G_off_fin}
\hG_{\sigma,3-\sigma}^B ({\bm \rho},{\bm \rho}', E) = \frac{r_{\sigma,2-\sigma}\bE}{r^4}
      (\hG_{22}^B + \hG_{00}^B - 2\hG_{11}^B).
\end{eqnarray}

Equations~(\ref{BL_G_Diag_fin}) and~(\ref{BL_G_off_fin}) are the final results for the
stationary Green function of an electron in bilayer graphene in the presence of
a magnetic field. The poles of $\hG^B({\bm \rho},{\bm \rho}', E)$
occur for $\bE_{ns}=0,\pm \sqrt{2}, \pm \sqrt{6},\ldots, \pm \sqrt{n(n-1)}$.
The residues of $\hG^B(E_{ns})$ can be obtained from~(\ref{BL_GDiag}) and~(\ref{BL_GOffDiag}).

In Figure~2 we plot the gauge-independent part of the dimensionless Green
function $\hbar\omega_c L^2\hG_{\sigma\sigma}^B({\bm \rho},0,E)$, calculated
for three values of~$\bE$. Similarly to electrons in monolayer graphene,
the components of the Green function $\hbar\omega_c L^2\hG^B({\bm \rho},0,E)$ decay exponentially for
large~$r$. For higher values of~$\bE=E/(\hbar\omega_c)$ they exhibit an oscillatory behavior.

Finally, we write the Green functions of monolayer and bilayer graphene taking into account contributions
from the inequivalent~$K'$ point of the Brillouin zone. For monolayer graphene the
Hamiltonian at the~$K'$ point is $\hat{H}^{M'}=(-\hat{H}^M)^T$ and its eigenstates are
$\psi_{nk_xs}^{M'}({\bm \rho})=e^{ik_xx}[\phi_n(\xi),s\phi_{n-1}(\xi)]/\sqrt{4\pi}$ \cite{Gusynin2007}.
In the basis $\{-s\phi^{K}_{n-1},\phi^{K}_{n},\phi^{K'}_n,s\phi^{K'}_{n-1}\}$, where the upper scrips
indicate the Brillouin zone point, the Green function is
\begin{equation}
\hG^{M}=\left(\begin{array}{cccc}
 \hG_{11}^M & -\hG_{10}^M &0 &0 \\ -\hG_{01}^M & \hG_{00}^M &0 &0 \\
 0 & 0 & \hG_{00}^M & \hG_{01}^M \\ 0 &0 &\hG_{10}^M & \hG_{11}^M
\end{array}\right).
\end{equation}

For bilayer graphene at the~$K'$ point the wave function is
$\psi_{nk_xs}^{B'}({\bm \rho})=e^{ik_xx}[\phi_n(\xi),-s\phi_{n-2}(\xi)]/\sqrt{4\pi}$ \cite{McCann2006}.
Thus the Green function of bilayer graphene in the
basis $\{-s\phi^{K}_{n-2},\phi^{K}_{n},\phi^{K'}_n,-s\phi^{K'}_{n-2}\}$ is
\begin{equation}
\hG^{B}=\left(\begin{array}{cccc}
 \hG_{22}^B & -\hG_{20}^B &0 &0 \\ -\hG_{02}^B & \hG_{00}^B &0 &0 \\
 0 & 0 & \hG_{00}^B & -\hG_{02}^B \\ 0 &0 &-\hG_{20}^B & \hG_{22}^B
\end{array}\right).
\end{equation}

To summarize, we calculated the Green functions for electrons in monolayer and bilayer graphene
in the presence of a magnetic field and expressed them in terms of the Whittaker functions.
The obtained formulas allow one to compute the Green functions using quickly converging expansions
of the Whittaker functions. This is a good starting point in more complicated calculations for
graphene monolayer and bilayer.

\hspace*{1em}

\end{document}